\documentclass[reprint,amsmath,amssymb,aps,prl,superscriptaddress]{revtex4-2}

\usepackage{graphicx}% Include figure files
\usepackage{dcolumn}% Align table columns on decimal point
\usepackage{bm}% bold math
\usepackage{hyperref}% add hypertext capabilities
\usepackage{overpic}
\usepackage{xcolor}
\usepackage{braket}
\usepackage{csquotes}
\newcommand{\panelwidth}{.95}
\usepackage{placeins}
\begin{document}

\title{Charging power enhancement at the phase transition\\
of a non-integrable quantum battery}
%\title{Many-Body Quantum Batteries: Do Integrability or Criticality\\ Enhance Charging Power?}
\author{D. Farina}
\email{donato.farina@unina.it}
\affiliation{Physics Department E. Pancini- Università degli Studi di Napoli Federico II,
Complesso Universitario Monte S. Angelo- Via Cintia- I-80126 Napoli, Italy}
\affiliation{INFN, Sezione di Napoli, Napoli, Italy}
\author{M. Sassetti}
\affiliation{Dipartimento di Fisica, Università di Genova, Via Dodecaneso 33, 16146 Genova, Italy}
\affiliation{CNR-SPIN, Via Dodecaneso 33, 16146 Genova, Italy}
\author{V. Cataudella}
\affiliation{Physics Department E. Pancini- Università degli Studi di Napoli Federico II,
Complesso Universitario Monte S. Angelo- Via Cintia- I-80126 Napoli, Italy}
\affiliation{CNR-SPIN, Via Dodecaneso 33, 16146 Genova, Italy}
\affiliation{INFN, Sezione di Napoli, Napoli, Italy}
\author{D. Ferraro}
\email{dario.ferraro@unige.it}
\affiliation{Dipartimento di Fisica, Università di Genova, Via Dodecaneso 33, 16146 Genova, Italy}
\affiliation{CNR-SPIN, Via Dodecaneso 33, 16146 Genova, Italy}
\author{N. Traverso Ziani}
\email{niccolo.traverso.ziani@unige.it}
\affiliation{Dipartimento di Fisica, Università di Genova, Via Dodecaneso 33, 16146 Genova, Italy}
\affiliation{CNR-SPIN, Via Dodecaneso 33, 16146 Genova, Italy}

\date{\today}

\begin{abstract}
\noindent
Exploiting many-body interaction and critical phenomena to improve the performance of quantum batteries is an emerging and promising line of research. A central question in this direction is whether quantum phase transitions can enhance the charging energy or the power. While preliminary works have addressed this problem in fine-tuned integrable models, its characterization in non-integrable systems remains limited due to the demanding numerical requirements. Here, we investigate a one-dimensional Axial Next-Nearest-Neighbor Ising model as an example of non-integrable quantum battery charged via a quantum-quench protocol. In contrast to integrable cases, we find that criticality in this setting can lead to a pronounced enhancement of the charging power.
Our findings inform quantum-battery design of many-qubit systems and are amenable to experimental verification on current quantum-simulation platforms, including neutral-atom arrays.
\end{abstract}

\maketitle

{\it Introduction.---}
Thermal phase transitions are a cornerstone of physics, as they are deeply connected to universality through the renormalization group~\cite{Goldenfeld1992}. At the same time, they are crucial for a plethora of indispensable technological applications, ranging from electric energy production to transportation~\cite{machines}. Quantum phase transitions—their zero‑temperature analogue—are equally fundamental for understanding complex quantum phenomena, such as magnetism and high‑temperature superconductivity~\cite{QPTs}. Yet their impact in quantum technologies remains largely unexplored. Notable exceptions include quantum thermal machines~\cite{Campisi2016, Bhattacharjee21}, quantum clocks~\cite{nello}, and, most relevant here, quantum batteries (QBs)~\cite{Quach23, RevModPhys.96.031001, Ferraro26}.

A QB is a quantum system designed to store and release energy on demand. These devices have emerged as a central topic in quantum thermodynamics, bridging condensed‑matter physics, quantum optics, and material science to explore how quantum effects can enhance energy storage and transfer ~\cite{RevModPhys.96.031001,Ferraro26, PhysRevE.87.042123,binder2015quantacell, Camposeo25}. This research effort spans foundational theoretical studies ~\cite{PhysRevLett.118.150601,PhysRevLett.128.140501,PhysRevLett.125.040601,PhysRevResearch.2.023095} and concrete experimental proposals ~\cite{PhysRevLett.120.117702, Quach22, Tibben25, Hymas25, Kurman26, Hu22, Gemme24, Joshi22, Cruz22}, with particular emphasis on many‑body implementations ~\cite{Rossini19, PhysRevA.97.022106,PhysRevLett.125.236402,rosa2020ultra}. Related concepts have also stimulated advances in other areas of quantum thermodynamics ~\cite{Campisi2016,PhysRevB.109.024310}.

Their appeal is related to two remarkable properties: They can show super-extensive charging, due to both collective and quantum effects~\cite{JuliaFarre20, Andolina19, Cavaliere25}, and they are expected to be more effective in exchanging energy with quantum devices due to a matching in time and energy scales~\cite{Crescente22}.

QBs are typically charged either through charger‑mediated protocols, where the charger itself is a quantum system ~\cite{PhysRevLett.120.117702,PhysRevB.98.205423,PhysRevB.99.035421}, or by means of a time dependent classical external drive~\cite{Zhang19}. In this second scenario, relevant for the present article, the analysis strongly benefits from the remarkable body of knowledge related to the physics of quantum quenches~\cite{mitra}. In this context, it has been natural to address integrable QBs. Here, short‑time charging power is expected to show no signatures of criticality~\cite{JuliaFarre20}, although critical points leave detectable imprints in the long‑time stored energy ~\cite{PhysRevLett.133.197001,grazi2025chargingFREE,grazi2025chargingXY,grazi2025universal}. This is due to the fact that integrability entails an extensive set of conserved quantities, which constrain dynamics and inhibits scrambling. Moreover, integrability is generically broken in realistic systems, since integrable models represent highly fine-tuned limits that are unstable to generic perturbations.

Motivated by this, we ask: can breaking integrability render quantum phase transitions relevant already at early times? This question is technologically significant, since, on the one hand, high-power quantum‑energy‑storage devices must operate on short timescales and, on the other hand, practical implementations rely on inherently non‑integrable architectures such as neutral‑atom arrays, Rydberg platforms, and trapped‑ion systems~\cite{Bruzewicz19, Weimer10, Browaeys20}.

In this work we provide a positive answer to this question. We consider the non‑integrable Axial Next‑Nearest‑Neighbor Ising (ANNNI) model~\cite{Selke88, Chakrabarti96}-the prime example of non-integrable spin chain with continuous phase transitions- as a QB, to show that criticality produces a clear enhancement of charging power, in sharp contrast to integrable cases.

\begin{figure}[h]
\centering
\begin{overpic}[width=.8\linewidth]{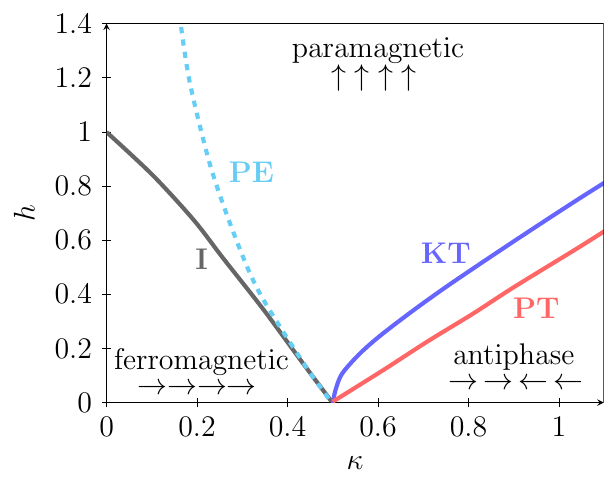}
\put(-5,60){\large (a)} 
\end{overpic}
\vspace{.3cm}

\begin{overpic}[width=.95\linewidth]{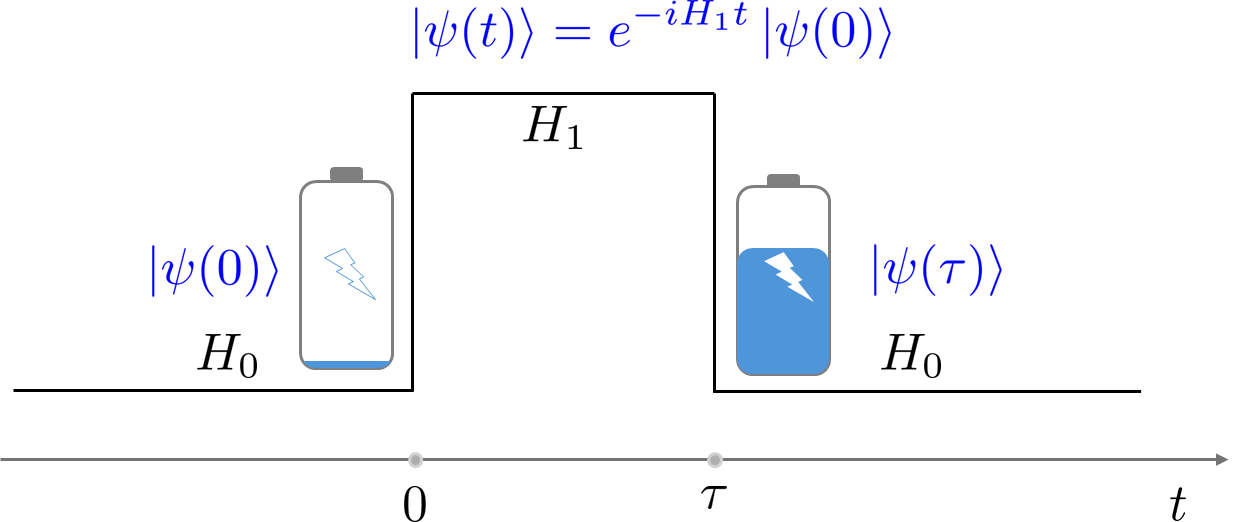}
\put(4,30){\large (b)} 
\end{overpic}
\caption{(a) Phase diagram of the ANNNI model in the $(\kappa,h)$ plane, adapted from Ref.\,\cite{cea2024exploring} ($J_1=1$).
The cyan dashed line is the 
exactly-solvable Peschel-Emery (PE) line. It partially overlap with the gray line representing the Ising phase transition (I).
We also report the 
Kosterlitz-Thouless
(blue line) and Pokrovsky-Talapov (red line) phase transitions (KT and PT respectively).
(b) Cartoon view of a QB charging protocol based on a double quantum quench of length $\tau$. Here, $|\psi(t) \rangle$ is the state of the QB at a given time $t$, $H_{0}$ indicates both the initial and the final Hamiltonian and $H_{1}$ the one used during the quench. The initial state $\ket{\psi(0)}$ is taken as the ground state of $H_0$.}
\label{fig:scheme}
\end{figure}
{\it ANNNI quantum battery.---}
We consider a spin QB based on the one dimensional ANNNI model~\cite{Selke88, Chakrabarti96} with Hamiltonian (assuming open boundary conditions)
\begin{equation}
    H=
- J_1 \sum_{i=1}^{L-1}  \sigma_i^x \sigma_{i+1}^x
- J_2  \sum_{i=1}^{L-2} \sigma_i^x \sigma_{i+2}^x
- h \sum_{i=1}^{L} \sigma_i^z\,.
\label{ANNNI}
\end{equation}
Here, $L$ is the number of sites, $\sigma^{x/z}_i$ are the Pauli matrices on the site $i$ in the usual representation, and $h$ is the external field. Moreover, $J_1$ and $J_2$ parametrize the strength of the spin-spin interactions, and lead to the frustration parameter
$
\kappa:=-J_2/J_1\,.
$
When $\kappa>0$ nearest neighbor and next-nearest neighbor interactions have opposite sign and hence compete.
We will focus on the zero temperature case, taking into account the fact that the reported results remain basically unaffected as long as the temperature is the smallest energy scale involved in the problem. 
Frustrated QBs~\cite{Zhang24, PRXQuantum.5.030319, catalano2024understanding, kovzic2025quantum, Bhattacharya26}, and in particular ANNNI-based implementations, have recently attracted increasing attention.
However, the influence of quantum phase transitions—and, more broadly, the role of criticality—on the charging power in this model has not yet been explored.

As shown in Fig.\,\ref{fig:scheme}(a), the ANNNI model features a rich quantum phase diagram~\cite{cea2024exploring} arising from the competition between nearest- and next-nearest-neighbor couplings. Specifically, it contains a ferromagnetic phase (where the nearest-neighbor interaction dominates), a paramagnetic phase (stabilized by a strong transverse field) and an antiphase configuration (where next-nearest-neighbor couplings prevail, though still influenced by the nearest-neighbor term). Between the paramagnetic and frustrated regimes lies the so-called floating phase, whose precise nature is still a subject of ongoing debate~\cite{Beccaria07}. In the Figure the dashed line marks the presence of an integrable Peschel-Emery line ~\cite{Peschel81, Wouters21}, while the boundaries among the different phases are indicated by continuous lines. Overall, we encounter the following
phase transitions by increasing $\kappa$:
Ising~\cite{QPTs} (gray), Kosterlitz-Thouless ~\cite{Kosterlitz73}
(blue), and Pokrovsky-Talapov~\cite{Pokrovsky79} (red).  All these lines meet at $h=0$ and $\kappa=0.5$. A peculiar feature of the present model, with important consequences for following analysis, is the fact that the integrable PE line and the Ising transition overlap only at small values of $h$ and then separate.

As a charging scheme, we adopt a double quantum-quench protocol, which we implement for the ANNNI model through numerical simulations. For completeness, we briefly summarize this protocol (see also Fig.\,\ref{fig:scheme}(b)). The ground state of the Hamiltonian $H_0$--the ANNNI model for the initial set of parameters--is assumed as the initial empty state of the QB, $\ket{\psi(0)}$. In symbols
$
H_0 \ket{\psi(0)} = E_{0} \ket{\psi(0)}
$, with $E_{0}$ the ground state energy. This choice is dictated by the simplicity of initializing the system in this state and by the fact that it is a passive state with respect to $H_0$~\cite{Pusz78}. It is then evolved via a step-constant time dependent Hamiltonian,
\begin{equation}
H(t) = 
\begin{cases}
H_1\,,\, \text{for\, } t \,\in\, (0,\tau)\\
H_0\,,\, \text{elsewhere} 
\end{cases}
\end{equation}
namely, the system is governed by $H_0$ at all times except during the charging window, where it evolves under $H_1$ for a controllable duration $\tau$. 
The unitary dynamics of the system is given by
\begin{equation}
\label{psit}
    \ket{\psi(t)} = e^{-i H_1 t} \ket{\psi(0)}\,,
\end{equation}
where the Hamiltonian $H_1$ is again of the ANNNI form shown in Eq. (\ref{ANNNI}), but with different values of the parameters with respect to $H_{0}$. The difference between the final (average) energy and the initial energy is the stored energy given by~\cite{RevModPhys.96.031001},
\begin{equation}
    W(\tau):=\bra{\psi(\tau)} H_0 \ket{\psi(\tau)} - E_{0}\,,
\end{equation}
which quantifies the energy injected via the protocol.
Accordingly, the (average) charging power is defined as  
$
P(\tau):=W(\tau)/\tau
$
which also characterize the speed of the energy transfer. 
The maximum power,
\begin{equation}
P_{\rm max}=\max_\tau P(\tau)\,,
\end{equation}
is the largest power achievable by appropriately tuning the charging time and will play a key role in the following analysis.
Finally, we note that, in the considered zero temperature limit and neglecting dissipative effects associated to environmental degrees of freedom, evolution is unitary and the system remains in a pure state at all times. Under this condition the notions of stored energy and maximum extractable work (\textit{ergotropy} ~\cite{allahverdyan2004maximal}) coincide~\cite{Farina19, Bhattacharjee21}.

%%%%%%%%%%%%%%%%%%%%%%%%%%%%%%

\begin{figure}
\vspace{.2cm}
\centering
\begin{overpic}[width=\panelwidth\linewidth]{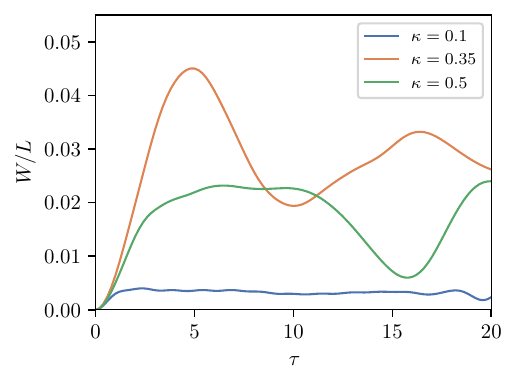}
\put(50,65){\large (a)}
\end{overpic}
\begin{overpic}[width=\panelwidth\linewidth]{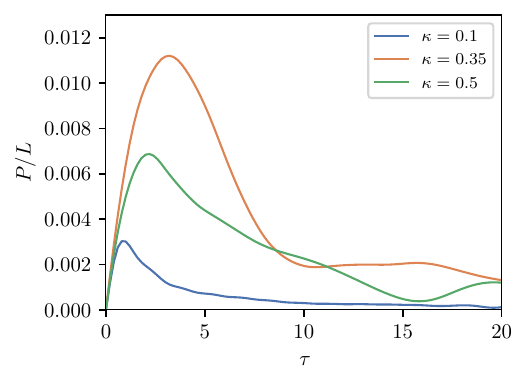}
\put(50,63){\large (b)} 
\end{overpic}
\caption{(a) Injected energy per spin as a function of the charging time $\tau$.
We consider $J_1=1$, $h=0.4$ for both $H_0$ and $H_1$ (open boundary conditions), plotting results for three different values of the frustration parameter $\kappa$ (see legend) and taking $\kappa'=\kappa+0.1$.
We report results for a chain length
$L=16$.
(b) Same as in (a) but for the charging power.
}
\label{fig:time-dependence}
\end{figure}

\begin{figure}
\vspace{.2cm}
\centering
\begin{overpic}[width=\panelwidth\linewidth]{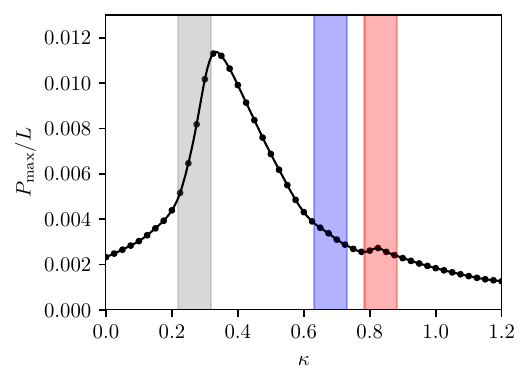}
\put(85,62){\large (a)}
\put(77,55){\normalsize $h=0.4$}
\end{overpic}
\begin{overpic}[width=\panelwidth\linewidth]{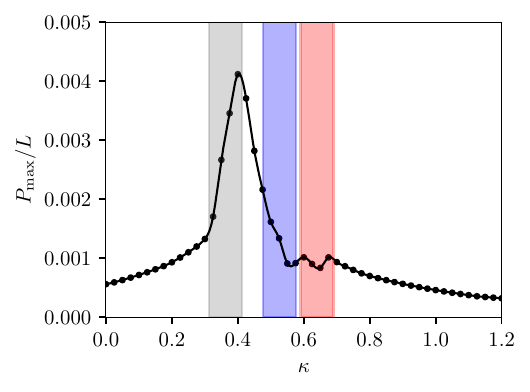}
\put(85,62){\large (b)} 
\put(77,55){\normalsize $h=0.2$}
\end{overpic}
\caption{(a) Maximum charging power per spin for a  16-spin chain with open boundary conditions, as a function of the frustration parameter $\kappa$ (with $\kappa'=\kappa+0.1$). 
The other parameters are set as $J_1=1$, $h=0.4$ for both $H_0$ and $H_1$. By increasing $\kappa$ we cross the following phase transitions:
Ising (gray), Kosterlitz-Thouless
(blue), and Pokrovsky-Talapov (red).
Points are calculated from exact diagonalization and interpolated via the black curve.
(b) Same as in (a) but for $h=0.2$.}
\label{fig:Pmax_16sites_obc_exact_0,2_0,4}
\end{figure}
{\it Results.---}%
As main focus of our analysis we start by considering genuinely non-integrable instances, taking the ANNNI model for both \(H_{0}\) and \(H_{1}\), and focusing on different regimes characterized by a fixed value of the transverse field \(h\). 
For sake of clarity, we adopt the convention that unprimed parameters ($h$, $\kappa$) refer
to the Hamiltonian $H_0$, while primed parameters ($h'$, $\kappa'$) refer to $H_1$. In
addition, we set $J_1 = 1$ in both Hamiltonians.
In the first considered protocol, the quantity
$\kappa$ sets the value of $J_2$ in $H_0$  and $\kappa'=\kappa+0.1$ sets the value of $J_2$ in $H_1$. 
Throughout the text, unless stated otherwise, we consider a 16-spin chain with open boundary conditions; data points are obtained via exact diagonalization and connected by the interpolating curves. We remark that the main results of the work remain qualitatively unchanged when considering larger systems. We show this explicitly in the Supplemental Material for a chain of $50$ spins studied using tensor network techniques.
\begin{figure*}
\centering
\begin{overpic}[width=.3\linewidth]{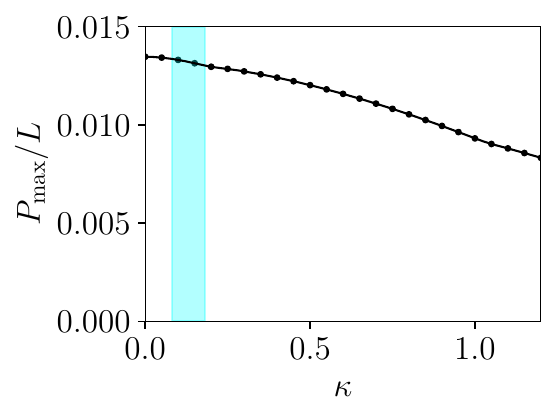}
\put(70,60){\large (a)} 
\put(50,30){\normalsize $h=1.2$}
\end{overpic}
\begin{overpic}[width=.3\linewidth]{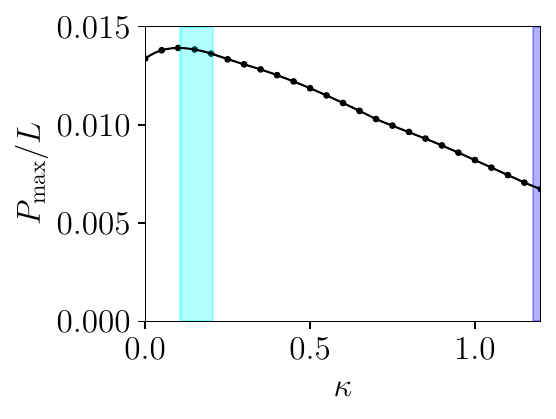}
\put(70,60){\large (b)} 
\put(50,30){\normalsize $h=1$}
\end{overpic}
\begin{overpic}[width=.3\linewidth]{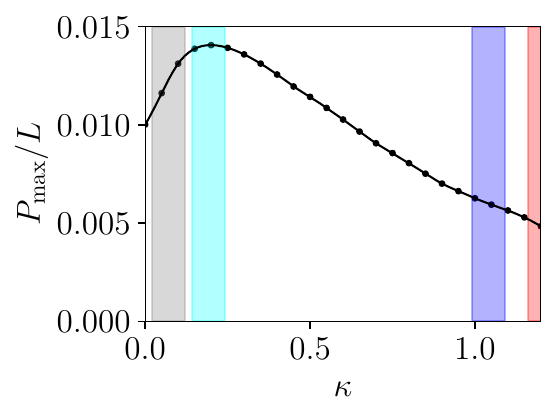}
\put(70,60){\large (c)} 
\put(50,30){\normalsize $h=0.8$}
\end{overpic}
\caption{%
(a)\,Maximum charging power per spin for a  16-spin chain with open boundary conditions, as a function of the frustration parameter $\kappa$ ($\kappa'=\kappa+0.1$). %The quantity $\kappa$ sets the value of $J_2$ in $H_0$  and $\kappa'=\kappa+0.1$ sets the value of $J_2$ in $H_1$. 
The other parameters are set as $J_1=1$, $h=1.2$ for both $H_0$ and $H_1$.
(b)\,Same as in (a) but for $h=1$.
(c)\,Same as in (a) but for $h=0.8$.
%
%Each vertical shadowed region starts when $H_1$ becomes critical and ends when $H_0$ becomes critical.
%
We have the following
phase transitions for increasing $\kappa$:
Ising (gray), Kosterlitz-Thouless
(blue), and Pokrovsky-Talapov (red).
We also report the region associated to the
exactly-solvable Peschel-Emery (cyan) line.
Points are calculated from exact diagonalization and interpolated via the black curve.}
\label{fig:Pmax_16sites_obc_exact_0,8_1_1,2}
\end{figure*}

\begin{figure}
\vspace{.2cm}
\centering
\begin{overpic}[width=\panelwidth\linewidth]{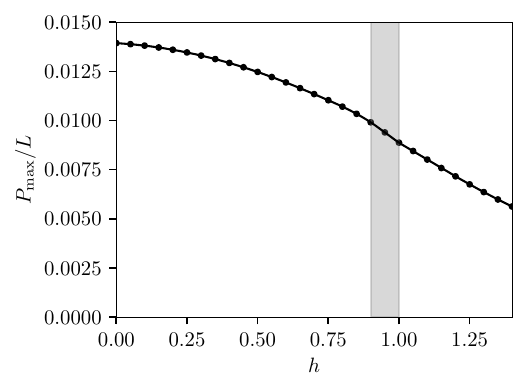}
\put(85,55){\large (a)}
\put(30,40){\normalsize TFI/TFI}
\end{overpic}
\begin{overpic}[width=\panelwidth\linewidth]{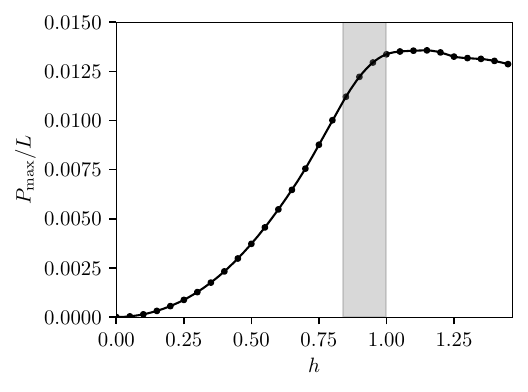}
\put(85,55){\large (b)} 
\put(30,40){\normalsize TFI/ANNNI}
\end{overpic}
\caption{(a)Maximum charging power per spin for a  16-spin chain with open boundary conditions, as a function of the field $h$ \textcolor{black}{and at fixed $\kappa=0$}. Both $H_0$ {(characterized by the transverse field $h$)} and $H_1$  {(characterized by the transverse field $h'=h+0.1$)}  realize the integrable Transverse Field Ising (TFI) model and no evident signature is present at the transition. 
Points are calculated from exact diagonalization and interpolated via the black curve.
(b) Same as in (a) but taking $h=h'$, $\kappa=0$ and $\kappa'=0.1$.
Here, $H_1$ realizes the non-integrable ANNNI model and the maximum power stabilizes at the Ising transition of $H_0$.
}
\label{fig:Pmax_16sites_obc_exact_tfi_stab}
\end{figure}

In Fig.\,\ref{fig:time-dependence} we show the time-dependence of the injected energy and the corresponding power, considering $h=0.4$ for both $H_0$ and $H_1$. Both quantities show a clear non-monotonic dependence on the frustration parameter $\kappa$ but, while the former shows a smooth oscillating behavior which is strongly suppressed at small values of $\kappa$, in the latter it is easy to identify a maximum. This maximum power per spin, $P_{\max}/L$, is numerically evaluated as the ordinate maximum for each curve in (b), and varying the frustration parameters. 

In the following plots colored shadowed regions will start when $H_1$ becomes critical and end when $H_0$ becomes critical.  Fig.\,\ref{fig:Pmax_16sites_obc_exact_0,2_0,4}\,(a) shows a clear peak in the maximum power per spin at the Ising transition of \(H_{0}\). This constitutes the main result of this work. This peaked structure in correspondence of the Ising phase transition becomes narrower for \(h = 0.2\), as shown in Fig.\,\ref{fig:Pmax_16sites_obc_exact_0,2_0,4}(b).
However, in this last case the overall charging power per spin decreases as we approach the line \(h = 0\), where \(H_{0}\) and \(H_{1}\) commute, implying that no energy can be injected.
Notice that less pronounced features are also present in correspondence to the other quantum phase transitions of the model.

To determine if the main feature is indeed related to the Ising phase transition or to the Peschel-Emery integrable line, which approximately coincide for the values of the transverse field considered above, we explore values of $h$ where the two depart. In Fig.\,\ref{fig:Pmax_16sites_obc_exact_0,8_1_1,2}\,(a) we set $h=1.2$. Here, we do not cross any critical line, although the system traverses the exactly solvable PE line; 
by itself, however, this does not produce any clear effect on the maximum power as testified by the overall decreasing but featureless curve of the maximum charging power as a function of $\kappa$. Fig.\,\ref{fig:Pmax_16sites_obc_exact_0,8_1_1,2}\,(b) (with $h=1$) shows that the Ising transition at \(\kappa = 0\) begins to noticeably modify the shape of the curve.
This effect becomes even more evident in Fig.\,\ref{fig:Pmax_16sites_obc_exact_0,8_1_1,2}(c) for \(h = 0.8\), where a clear peak emerges. 
%
%\vspace{.2cm}

We conclude our analysis by confirming that integrable models show no clear signatures of the phase transition in the maximum charging power. To do so in Fig.\,\ref{fig:Pmax_16sites_obc_exact_tfi_stab}\,(a)
the maximum charging power per spin is shown considering a charging protocol where the field $h$ is modified. The considered values are
$h$ and $h'=h+0.1$. The other parameters are set as $J_1=1$ and $J_2=0$ ($\kappa=0$), for both $H_0$ and $H_1$. Here, the integrable Transverse Field Ising (TFI) model is realized, and no clear signatures of the transition appear. 
We now consider a hybrid case in which \(H_{0}\) is a TFI Hamiltonian, while in \(H_{1}\) we activate the next-nearest-neighbor coupling with a sign opposite to that of the nearest-neighbor term.
The results are shown in Fig.\,\ref{fig:Pmax_16sites_obc_exact_tfi_stab}\,(b), where we take $h=h'$, $\kappa=0$ and $\kappa'=0.1$.
Here, \(H_{1}\) realizes the non-integrable ANNNI model, and the maximum power increases with $h$ until it stabilizes at the Ising transition of \(H_{0}\), yielding a striking manifestation of criticality.
%
%
%
%

%0.8 1 1,2 text%%%%%

{\it Discussion.---}
We have focused on the central question of whether quantum phase transitions can boost the performances of many-body quantum batteries. 
In contrast to integrable cases, we found that criticality in non-integrable settings can lead to a pronounced enhancements of the charging power.
Intuitively the reason behind this phenomenology is the following. Integrability entails an extensive set of conserved quantities, which constrain the dynamics and inhibit scrambling, whereas non-integrable models generally lack such constraints. Consequently, the dynamics is effectively faster, and the role of the equilibrium phase diagram becomes manifest at earlier times.
It should finally be noted that integrable models correspond to fine-tuned limits that occupy a measure-zero subset of the
many-body phase diagram. As a result, generic perturbations inevitably drive realistic
systems away from integrability, partly motivating the study of charging dynamics in
non-integrable regimes.
This highlights the essential role of quantum criticality and non-integrability in enabling fast and reliable quantum-battery operation by leveraging many-body effects.
Compelling directions for future work include analyzing the locally extractable work\, \cite{PhysRevA.107.012405,PhysRevLett.133.150402,PhysRevA.111.012212}, investigating non-Hermitian charging protocols\,\cite{el2018non,ashida2020non}, exploring the impact of dissipative dynamics\,\cite{PhysRevB.99.035421}, and consider- in higher dimensional models- the role of thermal criticality.
Our results inform quantum-battery design by clarifying how qubit interactions can be engineered to optimize charging performance, and
are amenable to experimental verification on current quantum-simulation platforms, including neutral-atom arrays.
%\vspace{.2cm}

{\it Acknowledgments.---} D. Farina acknowledges financial support from PNRR MUR Project No. PE0000023-NQSTI and from University of Catania via PNRR MUR Starting Grant project PE0000023-NQSTI.

%\clearpage

\FloatBarrier
\bibliography{sample}

\clearpage
\newpage

%{\large \bf End Matter}
%
%\vspace{.5cm}

%{\bf Appendix A: Further numerical results}
%\vspace{.3cm}
%

\begin{center}
{\Large \bf Supplemental Material 
% for \\
% \enquote{Charging power enhancement at the phase transition
% of a non-integrable quantum battery
% }
}
\end{center}

\section{Exact-Diagonalization Simulations}

In the main text, the charging dynamics is obtained from exact unitary evolution. 
Here we summarize the numerical
procedure used.
These calculations were implemented in Python
using standard scientific-computing libraries, including \textsc{NumPy} and
\textsc{SciPy}. 

%\subsection{Initial state and time evolution}
For each parameter set, we consider a spin-$1/2$ chain of length $L=16$ with open boundary
conditions. We construct two Hamiltonians, $H_0$ (used to define the stored energy) and
$H_1$ (used to generate the dynamics), both of ANNNI type, namely with nearest-neighbor and
next-nearest-neighbor $x$-$x$ couplings and transverse fields. 
The initial state $|\psi(0)\rangle$ is taken
as the ground state of $H_0$, obtained by sparse exact diagonalization. 
Starting from $|\psi(0)\rangle$, we compute the time-evolved
state under $H_1$,
via exact unitary evolution using sparse matrix-exponential methods.

%\subsection{Stored energy}
%
For a given charging time $\tau$, we evaluate the energy $
W(\tau)
$ injected with respect to $H_0$.
Technically, we sample the dynamics on a uniform time
grid $\tau\in[0,\tau_{\max}]$, considering $101$ points and $\tau_{\max}=20$, and export the energy
density $W(\tau)/L$.

%\subsection{Power and maximum power}
This is all we need to estimate numerically the average power per spin, computing $P(\tau)/L=W(\tau)/(\tau L)$, and its maximum value over time $P_{\max} /L=\max_{\tau} P(\tau)/L$.

\begin{figure}
\vspace{.2cm}
\centering
\begin{overpic}[width=\panelwidth\linewidth]{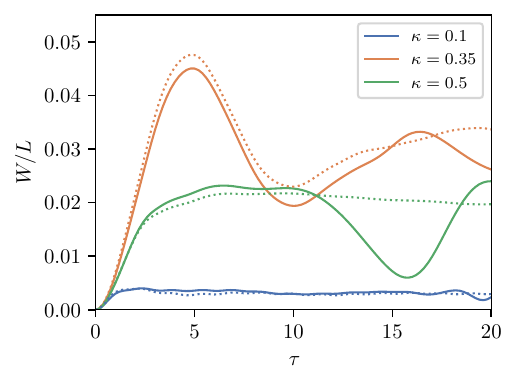}
\put(50,65){\large (a)}
\end{overpic}
\begin{overpic}[width=\panelwidth\linewidth]{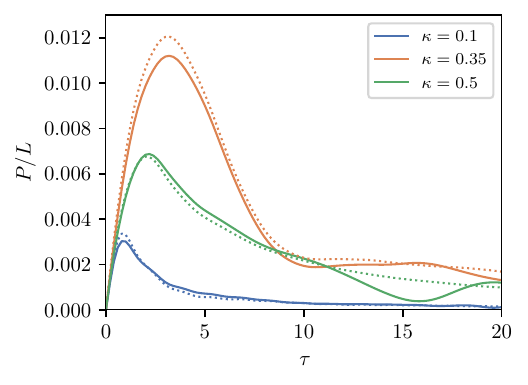}
\put(50,63){\large (b)} 
\end{overpic}
\caption{(a) Injected energy per spin as a function of the charging time $\tau$.
We consider $J_1=1$, $h=0.4$ for both $H_0$ and $H_1$ (open boundary conditions), plotting results for three different values of the frustration parameter $\kappa$ in $H_0$ (see legend) and taking $\kappa'=\kappa+0.1$ in $H_1$.
We report results for a chain length
$L=16$ using exact diagonalization (continuous curves, as in Fig.\,\ref{fig:time-dependence}) and $L=50$ 
using tensor-network estimations (dotted curves).
(b) Same as in (a) but for the charging power.
}
\label{fig:TN}
\end{figure}

\section{Tensor-Network Simulations}
To assess the robustness of our findings at larger sizes, we complement
those results here with tensor-network simulations based on matrix-product-state (MPS)
methods.  
Calculations were performed using the \textsc{TeNPy} Library \cite{tenpy2024}. 

We consider a one-dimensional spin chain of length $L=50$ with open boundary conditions.
The initial state is prepared as an approximate ground state of the Hamiltonian $H_0$
using the two-site density-matrix renormalization group (DMRG) algorithm, with maximum bond dimension $\chi=70$. 
Starting from
this state, the system is evolved in real time under a different Hamiltonian $H_1$ using
the single-site time-dependent variational principle (TDVP).

The time evolution is performed with a fixed time step $\Delta t = 0.2$. 
During the dynamics, we compute at each time step the expectation value
of the initial Hamiltonian $H_0$. 

Overall, within this scalable approach, we can efficiently estimate the quantity $W(\tau)$ for larger chain sizes 
and, through the usual postprocessing, compute the associated average power per spin and its maximum over time. 
%
%

% The resulting data show that the qualitative features observed in the exact
% diagonalization results (most notably the enhancement of the charging power in the
% vicinity of the Ising critical line) persist at larger system sizes, supporting the generality of the
% conclusions drawn in the main text.

A comparison between the results obtained from exact diagonalization and tensor-network
techniques is shown in Fig.\,\ref{fig:TN}. 
We consider the charging scheme already used for Figs.\,\ref{fig:time-dependence} and \ref{fig:Pmax_16sites_obc_exact_0,2_0,4}(a) of the main text, where we take constant jumps in the frustration parameters at fixed transverse field, specifically, $J_1=1$, $h=0.4$ in both $H_0$ and $H_1$, and $\kappa'=\kappa+0.1$. 
The results for $16$ 
spins (continuous curves) are the same as in Fig.\,\ref{fig:time-dependence}, showing the injected energy
and power per spin as functions of the charging time $\tau$. 
Notably, we observe that the
enhancement of the maximum charging power at the Ising transition persists when
considering larger system sizes ($50$ spins, dotted curves).
Qualitatively, this can be understood from the fact that the short-time dynamics is less
sensitive to the chain length.

\end{document}